\documentclass[twocolumn,showpacs,preprintnumbers,amsmath,amssymb]{revtex4}
\usepackage{graphicx}% Include figure files
\usepackage{dcolumn}% Align table columns on decimal point
\usepackage{bm}% bold math

\newcommand{\bq}    {\begin{equation}}
\newcommand{\eq}    {\end{equation}}
\newcommand{\bqr} {\begin{eqnarray}}
\newcommand{\eqr} {\end{eqnarray}}

\begin{document}
\title{Magnetic fields generated by r-modes in accreting millisecond pulsars}

\author{Carmine Cuofano}
\email{cuofano@fe.infn.it}
\author{Alessandro Drago}

\affiliation{Dipartimento di Fisica, Universit\'a di Ferrara 
and INFN sezione di Ferrara, 44100 Ferrara, Italy}

\begin{abstract}
In rotating neutron stars the existence of the Coriolis force allows
the presence of the so-called Rossby oscillations (r-modes) which are
known to be unstable to emission of gravitational waves.  Here, for the
first time, we introduce the magnetic damping rate in the evolution
equations of r-modes. We show that r-modes can generate very strong
toroidal fields in the core of accreting millisecond pulsars by
inducing differential rotation. We shortly discuss the instabilities
of the generated magnetic field and its long time-scale evolution in
order to clarify how the generated magnetic field can stabilize the star.
\end{abstract}


\maketitle
%
% \section{Introduction}
%
\noindent 
\section{Introduction}
R-mode oscillations are present in all rotating stars and they are
unstable to emission of gravitational waves
\cite{Andersson:2000mf}. These modes play therefore a very important
role in the astrophysics of compact stars and in the search for
gravitational waves.  R-mode instabilities are associated with
kinematical secular effects which generate differential rotation in
the star and large scale mass drifts, particularly in the azimuthal
direction. Differential rotation in turn can produce very strong
toroidal magnetic fields inside the star and these fields damp
instabilities converting the energy of the mode into magnetic energy.
This mechanism has been proposed in the case of rapidly rotating,
isolated and newly born neutron stars in
Refs.~\cite{Rezzolla:2001di,Rezzolla:2001dh}.  In our work we consider
the back-reaction of the magnetic fields on r-mode instabilities by
inserting for the first time the magnetic damping rate into the
evolution equations of r-modes.  In this way we can follow the
temporal evolution of both magnetic fields and r-modes even on a long
time scale.  In particular we show that taking into account the r-mode
instabilities it is possible to generate very strong magnetic fields
in accreting millisecond pulsars. \\ It is important to recall that
r-mode instabilities are damped also by the viscosity of the system,
either shear or bulk \cite{Andersson:2000mf}. At low temperature a
very important role could also be played by the so-called Ekman layer,
located at the interface between the solid crust and the fluid inner
core. The friction in the Ekman layer can be significantly enhanced
with respect to friction in a purely fluid component. Anyway it is very
difficult to give a precise estimate of this effect
\cite{Glampedakis:2006mn,Bondarescu:2007jw}, and in our analysis we
have not included the Ekman layer in order to study in a more clear
way the effects of magnetic fields on r-modes. We will anyway shortly
discuss how the presence of the Ekman layer can modify our results. \\
The paper is organized as follows: in Sec. II we write the r-mode equations
taking into account also the magnetic damping rate; in Sec. III we discuss
the equations regulating the time evolution of the magnetic field and of the
damping rate; in Sec. IV we discuss numerical solutions of the equations; 
in Sec. V we discuss open problems concerning the possible presence
of the Ekman layer, the effect of a superconducting shell and instabilities
of the generated internal magnetic field. Finally in Sec. VI we drive our conclusions.

\section{R-mode equations}
We want to derive the equations regulating the evolution of r-modes in
the presence of a pre-existent poloidal magnetic field $B_p$ and of
the generated internal field. We start by considering the conservation
of angular momentum, following Ref.~\cite{Wagoner:2002vr}.  The total
angular momentum $J$ of a star can be decomposed into an equilibrium
angular momentum $J_{*}$ and a canonical angular momentum $J_{c}$
proportional to the r-mode perturbation:
\begin{equation}
J=J_*(M,\Omega)+(1-K_j)J_c, \,\,\,\,\,\,\,\,\,\,\,\,\,  J_c=-K_c\alpha^2J_*
\label{eq1}
\end{equation}
where $K_{(j,c)}$ are dimensionless constants and
$J_{*}\cong I_{*}\Omega$. \\
Following Ref.~\cite{Friedman:1978hf} the canonical angular momentum
obeys the following equation:
\begin{eqnarray}
dJ_c/dt =&& 2J_c\{F_g(M,\Omega) \nonumber \\
&& -[F_v(M,\Omega,T_v)+F_{m_i}(M,\Omega,B)]\}
\label{eq2}
\end{eqnarray}
where $F_g$ is the gravitational radiation growth rate, $F_v=F_s+F_b$ 
is the sum of the shear and bulk viscous damping
rate, $T_v(t)$ is a spatially averaged temperature and now we also have
introduced the damping rate $F_{m_i}$, associated with the
generated internal magnetic field. The explicit expression of
$F_{m_i}$ will be discussed in the following. \\
The total angular momentum satisfies the equation:
\begin{equation}
dJ/dt=2J_c F_g+\dot{J}_a(t)-I_{*}\Omega F_{m_e}
\label{eq3}
\end{equation}
where $\dot{J}_a$ is the rate of accretion of angular momentum (we
have assumed it to be $\dot{J}_a=\dot{M}(GMR)^{1/2}$, see
Ref.~\cite{Andersson:2001ev}) and $F_{m_e}$ is the magnetic breaking
rate associated to the external poloidal magnetic field. In the
present paper we are not considering further breaking mechanisms as
e.g. the interaction between the magnetic field and the accretion
disk.  Combining Eqs.~(\ref{eq2}) and (\ref{eq3}) we obtain the
evolution equations of the r-mode amplitude $\alpha$ and of the
angular velocity of the star $\Omega$:
\begin{eqnarray}
\frac{d\alpha}{dt} =&& \alpha(F_g-F_v-F_{m_i}) \nonumber \\
&& +\alpha[K_jF_g+(1-K_j)(F_v+F_{m_i})]K_c\alpha^2 \nonumber \\
&& -\frac{\alpha\dot{M}}{2\tilde{I}\Omega}
\left(\frac{G}{MR^3}\right)^{1/2}+\frac{\alpha F_{m_e}}{2} 
\label{eq4a} \\
\frac{d\Omega}{dt} =&& -2K_c\Omega\alpha^{2}
[K_jF_g+(1-K_j)(F_v+F_{m_i})] \nonumber \\
&& -\frac{\dot{M}\Omega}{M} +\frac{\dot{M}}{\tilde{I}}
\left(\frac{G}{MR^3}\right)^{1/2}
-\Omega F_{m_e} \;
\label{eq4b}
\end{eqnarray}
where $I_*=\tilde{I}MR^2$ with $\tilde{I}=0.261$ for an n=1 polytrope
and $K_c=9.4\times 10^{-2}$, see Ref.~\cite{Owen:1998xg}. Our results
turn out to be rather insensitive to the value of $K_j \sim 1$ (see
Ref.~\cite{Wagoner:2002vr}).  \\
\begin{figure}
\begin{center}
\includegraphics[height=8cm, width=9cm]{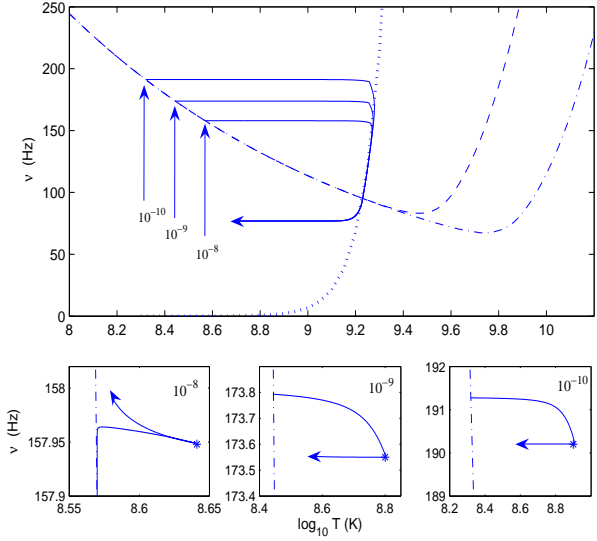}
\end{center}
\caption{\label{fig1} Top panel: instability region in the Temperature vs
Frequency plane obtained using either the
bulk viscosity damping rate $F_b$ given in \cite{Andersson:2000mf} 
(region above the dot-dashed line) or the $F_b$ given in \cite{Owen:1998xg} (region above the dashed line).
Also shown are the paths followed by accreting neutron stars, 
without toroidal magnetic fields (solid lines). The three curves 
correspond to different values of accretion rate 
$\dot{M}=(10^{-8},10^{-9},10^{-10})\,\,\mbox{M}_{\odot}\,\,\mbox{yr}^{-1}$.  
We plot also the temperature equilibrium curve
\cite{Bondarescu:2007jw} (dotted line), obtained taking into account
the neutrino cooling and the reheating due to viscosity. 
Bottom panels: new paths obtained taking into account the new generated toroidal
fields. Here $B_d=10^8$~G. The moments at which
toroidal magnetic fields damp r-mode instabilities are indicated by
asterisks.}
\end{figure}
\section{Magnetic damping}
The crucial ingredient introduced before is the magnetic
damping rate $F_{m_i}$, which we have inserted in the evolution of r-modes.
The expression of the magnetic damping rate has been derived 
in \cite{Rezzolla:2001di,Rezzolla:2001dh}, where it has been shown that
while the star remains in the instability region, the r-modes generate
a differential rotation which can greatly amplify a pre-existent magnetic
field. 
More specifically, if a poloidal magnetic field was originally present,
a strong toroidal field is generated inside the star.
The energy of the modes is therefore transferred to the magnetic field and
the instability is damped.
\\
We assume that the stellar magnetic field $\textbf{B}$ is initially
dipolar and aligned with the star's spin axis
\begin{eqnarray}
\mbox{\textbf{B}}_0=\mbox{\textbf{B}}^{p}(t=0)=B_d \frac{R^3}{r^3}(2
\, \texttt{cos} \, \theta \, \mbox{\textbf{e}}_r+ \, \texttt{sin} \,
\theta \, \mbox{\textbf{e}}_\theta)
\label{eq1M}
\end{eqnarray}
where $B_d$ is the strength of the equatorial magnetic field at the
stellar surface. The initial magnetic field is assumed to be described
by Eq.~(\ref{eq1M}) both inside and outside. 
In order to avoid divergences we assume the previous radial
dependence of the dipolar field to hold only for $r \geq R/2$.
If for instance the dipolar field remains roughly constant in the inner region
its contribution to the total energy of the final toroidal field is negligible. 
Therefore in the following and in agreement with Ref.~\cite{Rezzolla:2001dh}, 
radial integrations extend from $R/2$ to $R$. \\
To estimate the magnetic field produced by r-modes we start by
writing the $l=m=2$ contribution to the perturbation velocity:
\begin{eqnarray}
\delta\textbf{v}(r,\theta,\phi,t)=\alpha\Omega
R\left(\frac{r}{R}\right)^2\textbf{Y}_{22}^Be^{i\sigma t}.
\end{eqnarray}
Following Ref.~\cite{Rezzolla:2001di} we get the total azimuthal displacement
from the onset of the oscillation at $t_0$ up to time $t$, which reads:
\begin{eqnarray}
&&\Delta \tilde{x}^{\phi}(r,t)\equiv \int_{t_0}^t \delta v^\phi(t')dt'  \nonumber \\
&& = \frac{2}{3}\left(\frac{r}{R}\right)
k_2(\theta)\int_{t_0}^t\alpha^2(t')\Omega(t')dt'+\mathcal{O}(\alpha^3) \,\,\,\,\,\,\,\,\,
\end{eqnarray}
where $k_2(\theta)\equiv (1/2)^7(5!/\pi)(\texttt{sin}^2\theta-2\texttt{cos}^2\theta)$.
The relation between the new and the original magnetic field inside the star 
in the Lagrangian approach reads~\cite{Rezzolla:2001di}:
\begin{eqnarray}
\frac{B^j}{\rho}(\tilde{\textbf{x}},t)=\frac{B^k}{\rho}(\textbf{x},t_0)
\frac{\partial\tilde{x_j}(t)}{\partial x^k(t_0)}.
\end{eqnarray}
This equation implies that the radial dependence of the initial and 
final magnetic field is the same.
Integrating on time the induction equation in the Eulerian approach one gets~\cite{Rezzolla:2001dh}:
\begin{eqnarray}
\delta B^{\theta} &\simeq& \delta B^r \simeq 0  \nonumber \\
\delta B^{\phi} &\simeq& B_0^{\theta} \int \dot{\phi}(t') dt' 
\simeq B_0^{\theta}\int\frac{\delta v^\phi(t')}{r}dt'
\label{eq2M}
\end{eqnarray}
where $B_{\phi}$ is the toroidal component.\\
The expression of the magnetic damping rate reads:
\begin{eqnarray}
F_{m_i} (t) &&= \frac{(dE_M/dt)}{\tilde{E}} \nonumber \\ 
\simeq && \frac{4(1-p)}{9\pi p \cdot (8.2\times 10^{-3})}
\frac{B_p^2 R \Lambda '\int^t_0 \alpha^2(t ')\Omega(t ') dt '}{M\Omega}\,\,\,\,\,\,\,\,\,\,\,
\label{eq14}
\end{eqnarray}
where $\tilde{E}$ is the energy of the mode, $E_M$ is the magnetic
energy, $\Lambda '\approx\mathcal{O}(1)$ is a dimensionless parameter \cite{Rezzolla:2001dh}
and $p=0.5$ indicates that we are integrating from $R/2$ to $R$. 
The time integral over the r-mode amplitude $\alpha$ takes contribution from the period
during which the star is inside the instability region.  
\section{Numerical solutions}
In our numerical analysis we use the estimate given in
\cite{Andersson:2000mf} for the gravitational radiation reaction rate
due to the $l = m = 2$ current multipole
\begin{equation}
F_{g} = \frac{1}{47}M_{1.4}R_{10}^{4}P^{-6}_{-3} \,\,\,\,\, \mbox{s}^{-1}
\label{eq0A}
\end{equation}
as well as for the bulk and shear viscosity damping rates
\begin{eqnarray}
F_{b} &=& \frac{1}{2.7\times 10^{11}} M^{-1}_{1.4}R_{10}
P^{-2}_{-3}T^{6}_{9} \,\,\,\,\, \mbox{s}^{-1} \nonumber \\
F_{s} &=& \frac{1}{2.2\times 10^{7}}
M_{1.4}R_{10}^{-5}T^{-2}_{9} \,\,\,\,\, \mbox{s}^{-1}
\label{eq0B}
\end{eqnarray}
where we have used the notation $M_{1.4}=M/1.4 \,\, M_{\odot}$,
$R_{10}=R/10$ Km, $P_{-3}=P/1$ ms and $T_9=T/10^9$ K. 
It is important to remark that the damping rate $F_b$ 
depends on the density profile of the star. Using a different
profile in \cite{Owen:1998xg} they obtained a value for $F_b$ 
three orders of magnitude larger than in \cite{Andersson:2000mf}.
Fortunately our results are insensitive to this quantity, only
the instability window is modified as shown in Fig.~\ref{fig1}.
\\ 
Viscosity depends critically on temperature. We include three factors in
modeling the temperature evolution: modified Urca cooling, shear
viscosity reheating, and accretion heating.  The cooling rate due to
the modified Urca reactions, $\dot{\epsilon}_{u}$, reads
\cite{Shapiro83}
\begin{equation}
 \dot{\epsilon}_{u}=7.5\times 10^{39}M_{1.4}^{2/3}T_{9}^{8} 
\,\,\,\,\,\, \mbox{erg s}^{-1} \,.
\end{equation}
The neutron star will be heated by the action of shear viscosity on the
r-mode oscillations. The heating rate due to shear viscosity, 
$\dot{\epsilon}_{s}$,
reads \cite{Andersson:2000mf}
\begin{eqnarray}
 \dot{\epsilon}_s &=& 2\alpha^2\Omega^2MR^2\tilde{J}F_s \nonumber\\
 &=& 8.3\times10^{37}\alpha^2\Omega^2\tilde{J}M_{1.4}^{9/4}
 R_{10}^{-15/4}T_{9}^{-2} \,\,\,\,\,\, \mbox{erg s}^{-1} \,. \,\,\,\,\,\,\,\,
\end{eqnarray}
where $\tilde{J}=1.635\times 10^{-2}$. \\
Accretion heating has two components. We use the estimates given in 
\cite{Watts:2001ej}.
The first contribution arises when accreted matter undergoes nuclear 
burning at the surface of the star
\begin{equation}
 \dot{\epsilon}_n=\frac{\dot{M}}{m_{B}}\times 1.5 \,\, \mbox{MeV} = 4\times 10^{51} 
 \dot{M}_{1.4} \,\,\,\,\, \mbox{erg s}^{-1}
\label{eq10}
\end{equation}
where $m_B$ is the mass of a baryon. \\
The second contribution arises because the flow is assumed to be
advection dominated. 
The heating rate is
\begin{equation}
 \dot{\epsilon}_h \sim \frac{R}{\lambda}\frac{GM\dot{M}}{R} 
 = 8\times 10^{51} M_{1.4}^{13/6}\dot{M}_{1.4} \,\,\,\,\, \mbox{erg s}^{-1} \,.
\label{eq11}
\end{equation}
Finally we use the estimate of the heat capacity $C_v$ given in 
\cite{Watts:2001ej}
\begin{equation}
 C_v=1.6\times 10^{39}M_{1.4}^{1/3}T_{9} \,\,\,\,\, \mbox{erg K}^{-1}.
\label{eq12}
\end{equation}
The equation of thermal balance of the star is
\begin{equation}
 \frac{d}{dt}\left[\frac{1}{2}C_{v}T \right]= \dot{\epsilon}_s
-\dot{\epsilon}_u+\dot{\epsilon}_n+\dot{\epsilon}_h \,\, .
\label{eq13}
\end{equation}
We consider a scenario in which the mass accretion spins an initially slowly
rotating neutron star up to millisecond period and we investigate the evolution of
internal toroidal magnetic fields when the star enters the r-modes 
instability window.
\begin{figure}
\begin{center}
\includegraphics[height=8cm, width=9cm]{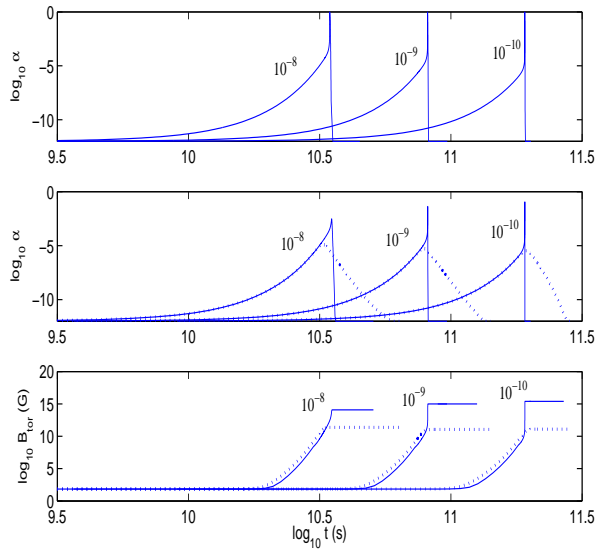}
\end{center}
\caption{\label{fig2} Top panel: temporal evolution of r-modes 
amplitude without toroidal fields.
Middle panel: temporal evolution of r-modes amplitude with toroidal fields for
$B_d=10^8$ G (solid line) and $B_d=10^{9}$ G (dotted line). Bottom
panel: temporal evolution of volume averaged toroidal magnetic fields. 
We consider 
$\dot{M}=(10^{-8},10^{-9},10^{-10})\,\,\mbox{M}_{\odot}\,\,\mbox{yr}^{-1}$.}
\end{figure}
\\
We start by discussing the evolution of temperature and spin frequency
obtained without magnetic fields. In Fig.~\ref{fig1} we show that the
star crosses the r-mode instability region in a regime of
thermo-gravitational runaway \cite{Bondarescu:2007jw}.  R-modes grow
exponentially due to the decrease of the shear viscosity with
increasing temperature.  As a consequence r-mode amplitudes rapidly
reach the saturation value (we chose $\alpha_{sat}=1$ although, as we
shall see later, magnetic fields limit $\alpha\ll\alpha_{sat}$) and
the viscosity heats significantly the star.  At this stage the star
loses angular momentum by emission of gravitational waves and goes out
of the instability region in a time of hundred of years.
\\
Taking into account magnetic fields, the evolutionary scenario for the
star is quite different.  We discuss results obtained solving in a self-consistent
way Eqs.~(\ref{eq4a},\ref{eq4b},\ref{eq14},\ref{eq13}). From a practical viewpoint
we proceed as follows: we first solve Eqs.~(\ref{eq4a},\ref{eq4b},\ref{eq13}) and we get
an estimate of $F_{m_i}(t)$ inserting the results in Eq.~(\ref{eq14}); we use this estimate to 
solve again Eqs.~(\ref{eq4a},\ref{eq4b},\ref{eq13}). This procedure is iterated till
numerical convergence is reached. \\
The exponential growth of r-modes induces extremely large secular
effects and the toroidal magnetic field is either produced or
amplified by the wrapping of the poloidal field produced by the
secular velocity field which is mostly toroidal. In the bottom panel of Fig.~\ref{fig1} we show 
the new trajectory of the star in the Temperature-Frequency plane obtained 
taking into account the generated toroidal fields, and we indicate with asterisks 
the moments at which magnetic fields damp r-modes.  It is important to remark that this
happens when the star is still in the region which was unstable taking into account only the
viscous damping.   \\ 
In Fig.~\ref{fig2} we show the
evolution of the r-mode amplitude $\alpha$, without magnetic field
(top panel) and with magnetic field (middle panel). Three different
values of accretion rate $\dot{M}$ and two values of the initial 
poloidal magnetic field $B_d$ are considered. 
In the scenario with magnetic fields, the maximum values
of $\alpha$ are in the range $\alpha_{max}\sim[10^{-6}-10^{-1}]$ and
the generated toroidal fields are in the range
$B_{tor}\sim[10^{11}-10^{15}]$~G (Fig.~\ref{fig2} bottom panel).  In
this analysis we have considered values of accretion rate $\dot{M}$ and
magnetic field $B_d$ typical of accreting Low Mass X-Ray binaries
(LMXBs).
\begin{figure}
\begin{center}
\includegraphics[height=7cm, width=9cm]{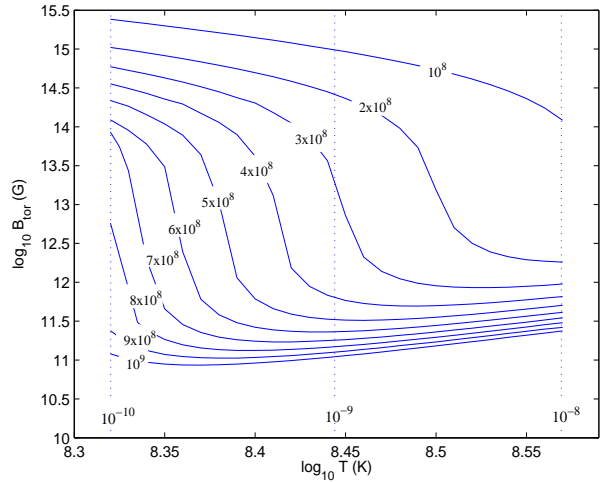}
\end{center}
\caption{\label{fig3} Volume averaged toroidal magnetic field generated by r-modes as
 a function of the temperature at which the star enters the
 instability window. For reference we show also the value of the temperatures 
 which correspond to a  mass accretion rate
 $\dot{M}=(10^{-8},10^{-9},10^{-10})\,\,\mbox{M}_{\odot}\,\,\mbox{yr}^{-1}$.
 The various curves correspond to different values of the initial poloidal
 magnetic field, in the range $[10^8-10^9]$ G.}
\end{figure}
\\
In Fig.~\ref{fig3} we show the generated toroidal magnetic field as a
function of the temperature at which the star enters the instability
window. We display the volume averaged field obtained at the moment in which
the field itself has completely damped r-modes. The value of the magnetic field
is the asymptotic one displayed in the bottom panel of Fig.~\ref{fig2}.
Results of Fig.~\ref{fig3} are obtained using Eqs.~(\ref{eq10},\ref{eq11}) 
to describe the connection between accretion rate $\dot{M}$ and temperature $T$, but
they are rather insensitive to the exact relation between these two quantities.  
\section{Open problems}
\begin{figure}
\begin{center}
\includegraphics[width=9cm,height=6cm]{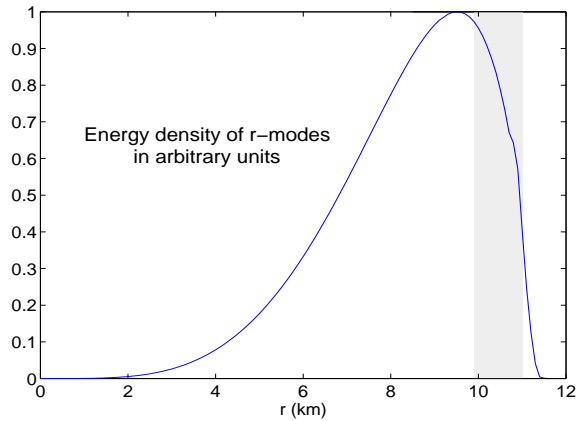}
\end{center}
\caption{\label{fig4} Energy density of r-modes as a function of the star radius.
The estimates about the size of the superconducting region are quite uncertain. 
Recent results obtained in \cite{Baldo:2007} suggest that the superconducting shell
has a thickness $\lesssim 1$ km and it is located in the grey area.}
\end{figure}

In our work we have not discussed the possible existence of Ekman
layer. If this layer is present it stabilizes the star up to
frequencies of a few hundred Hertz. The scenario discussed above does
not change qualitatively, but since the star enters the instability
region at higher frequencies, the growth rate of magnetic fields is
larger.  \\
Another important open point concerns the possible formation of superconductivity 
in the core of neutron stars. A fraction of the core, whose temperature is below the
critical value $T_c \sim~10^9$ K, is expected to be either a 
Type I \cite{Link:2003, Buckley:2004} or a Type II superconductor \cite{Baldo:2007,Wasserman:2003}. 
The exact nature of the possible superconducting layer is still uncertain 
as well as the precise value of the superconducting gap.
The fundamental quantity in our analysis is the size of the superconducting region
which we show in Fig.~\ref{fig4} using the results obtained in \cite{Baldo:2007}. The thickness
of the superconducting shell is about $1$~km.
In the same figure we also display the energy density of the r-modes. It is clear
that most of the volume where r-modes can develop remains not superconducting
and in that region our analysis can still be applied: r-modes are damped 
by the magnetic field and therefore the perturbation velocity drops to zero. 
R-modes can develop in the thin superconducting layer, but their amplitude is strongly suppressed
because the perturbation velocity has to vanish at the interface between the
superconducting and the not superconducting region. Moreover, if the superconductivity
is of Type II, the magnetic flux of the original poloidal field organizes
into quantized flux tubes. R-modes will stretch these flux tubes,
increasing their length and thus changing their magnetic energy. 
As in the case of the normal core, this process can generate very intense
magnetic fields which lead to the damping of the r-mode oscillations
even in the superconducting layer \cite{Rezzolla:2001dh}. \\
Several issues remain open concerning how the new generated magnetic
fields are affected by possible instabilities.  In the stably
stratified environment of a stellar interior there are two types of
instabilities: the Parker (or magnetic buoyancy) and the Tayler
instabilities (or pinch-type), both driven by the magnetic field
energy in the toroidal field.  The buoyancy instability is negligible
for $B_{tor}\lesssim 10^{15}$~G, so we focus on Tayler instabilities
because they set in at a lower field strength \cite{Spruit:1999cc}.  It
is important to remark that in a stable and stratified neutron star
the condition for the Tayler instability reads \cite{Spruit:2001tz}:
\begin{eqnarray}
\frac{\omega_{A}}{\Omega}> \left(\frac{N_{\mu}}{\Omega}\right)^{1/2}
\left(\frac{\eta}{r^2\Omega}\right)^{1/4}
\label{Eqtayler}
\end{eqnarray}
where $\omega_{A}=B/(4\pi\rho)^{1/2}r$ is the Alfv\'en frequency,
$N_{\mu}\simeq 5\times10^4 \,\,\mbox{s}^{-1}$ is the compositional
contribution to the buoyancy frequency and $\eta\sim10^{-9}\,\,
\mbox{cm}^2\,\mbox{s}^{-1}$ is the magnetic diffusivity \cite{Haensel:2007yy}.
From Eq.~(\ref{Eqtayler}) we can conclude that 
in the stably stratified core of a neutron star, the
Tayler instability sets in for $B_{tor}^{cr}\gtrsim 10^{12}$~G. \\
After the development of the Tayler instability, the toroidal
component of the field produces, as a result of its decay, a new
poloidal component which can then be wound up itself, closing the
dynamo loop. Both components then grow, more slowly, until the
saturation level is reached, when the field is being destroyed by the
instability at the same rate at which it is being amplified by the
differential rotation \cite{Braithwaite:2005pa}.  When the
differential rotation stops the field can evolve into a stable
configuration of a mixed poloidal-toroidal twisted-torus shape
embedded inside the star with an approximately dipolar field connected
to it outside the star
\cite{Reisenegger:2008yk,Braithwaite:2005md,Braithwaite:2005ps,Braithwaite:2005xi}.
Once the field is stabilized it should evolve as a result of diffusive
processes as Ohmic dissipation, ambipolar diffusion, and Hall drift
\cite{Goldreich:1992ap}, whose typical time-scales are:
\begin{eqnarray}
 t_{ohmic} &\sim& 2\times 10^{11} \, \frac{L_{5}^{2}}{T_{8}^{2}}
 \left(\frac{\rho}{\rho_{nuc}}\right)^{3} \,\, \mbox{yr} \label{eq21} \\
 t_{ambip} &\sim& 3\times 10^{9} \, \frac{T_8^2 L_5^2}{B_{12}^2} 
\,\, \mbox{yr} \label{eq22} \\
t_{Hall} &\sim& 5\times 10^{8} \, \frac{L_{5}^2}{B_{12}}
\left(\frac{\rho}{\rho_{nuc}}\right) \,\, \mbox{yr}
 \label{eq23}
\end{eqnarray}
where $L_5 = L/10^5 $~cm is the size of the region embedding the
magnetic field.
\section{Conclusions}
We have shown how r-modes can generate strong toroidal
fields in the core of accreting millisecond neutron stars, and how
these fields influence the growth rate of r-mode instabilities. 
\begin{figure}
\begin{center}
\includegraphics[height=7cm, width=9cm]{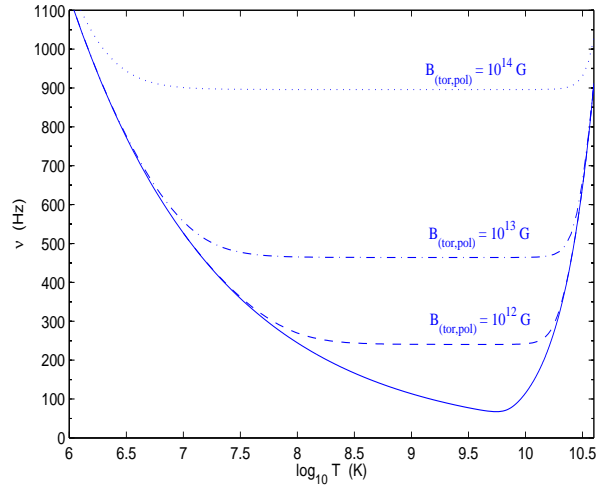}
\end{center}
\caption{\label{fig5} New r-modes instability regions for neutron stars with stable configurations
of mixed poloidal-toroidal fields in the inner core (no magnetic field - solid line; 
$B_{tor,pol}=10^{12}$~G - dashed line; $B_{tor,pol}=10^{13}$~G - dotted-dashed line;
$B_{tor,pol}=10^{14}$~G - dotted line). We assume that poloidal and toroidal components
have similar strengths.}
\end{figure}
Tayler instability sets in for strengths of the generated fields of
the order of $10^{12}$~G and stabilizes the toroidal component
by producing a new poloidal field of similar strength. This stable
configuration evolves on a time-scale, regulated by diffusive
processes. Our results imply that in the core of accreting neutron stars in LMXBs,
rotating at frequencies $\nu\gtrsim 200$~Hz, there are strong magnetic
fields with strengths $B\gtrsim 10^{12}$~G. 
\\
Finally, it is tempting to try to investigate how the new stable configuration
of magnetic fields modifies the instability window of r-modes.
More explicitly the scenario we have in mind is the following:
\begin{itemize}

 \item the compact star inside the LMXB initially enters the instability window
 and it follows the trajectories that are displayed in the bottom panel 
 of Fig.~\ref{fig1} and in Fig.~\ref{fig2}. When the toroidal field reaches 
 the critical value dictated by the Tayler instability, the poloidal field 
 suddenly increases and we assume it reaches a value of the same order of $B_{tor}^{cr}$.
 The new magnetic configuration, in which the poloidal and
 the toroidal field are of the same order of magnitude, is stable.
 In the new configuration the internal poloidal field is a few orders of magnitude larger
 than the initial poloidal field. R-modes try again to deform the poloidal field 
 generating a new toroidal component but the magnetic damping rate given by Eq.~(\ref{eq14})
 is now much larger and the star is stable with respect to r-modes up to frequencies
 of the order of $200$~Hz.

 \item the star continues accreting and increasing its frequency and it enters again
 the instability region generating again a new toroidal field.
 When the new toroidal component becomes much larger than the poloidal component
 Tayler instability sets in again increasing the value of the poloidal field.
 In Fig.~\ref{fig5} we show how the instability window changes in dependence of the 
 internal magnetic configuration, assuming that the toroidal and poloidal components
 are equal.
 
 \item to determine the actual maximum value of the limiting frequency we take into account
 also the time-scale associated with the diffusive processes described by 
 Eqs.(\ref{eq21}--\ref{eq23}). If the magnetic fields exceed a few $10^{14}$~G the diffusion
 time-scale becomes shorter than $10^6$~yr and the star cannot accelerate further even for very
 large values of $\dot{M}$.

\end{itemize}

It is clear that even in the absence of the Ekman
layer the new internal magnetic fields can stabilize stars with frequencies up to several hundreds Hertz.
The possible presence of a superconducting shell would screen these internal fields so that they
would not affect the dynamics of the external region.
The phenomenological implications of this model will be discussed in a forthcoming paper.
\newline

It is a pleasure to thank Marcello Baldo, Giuseppe Pagliara and Andreas Reisenegger 
for many useful and stimulating discussions.

\bibliography{bibliografia}

\end{document}